\documentclass{aa}
\usepackage{graphicx}

\begin{document}
\title{An inverse Compton scattering (ICS) model of pulsar emission:
II. frequency behavior of pulse profiles}
\author{G. J. Qiao\inst{1,2}, J. F. Liu\inst{1,2}, B. Zhang\inst{1,4},
 J. L. Han\inst{2,3}}
\offprints{G. J. Qiao.\\ gjn@pku.edu.cn}
\institute{Astronomy Department, Peking University, Beijing
100871, China
        \and Chinese Academy of Science - Peking University
joint Beijing Astrophysics Center, Beijing 100871, China
        \and National Astronomical
Observatories, Chinese Academy of Sciences, Beijing 100012, China
    \and Present address: Astronomy \& Astrophysics Department,
         Pennsylvania State University, University Park, PA 16802, USA }
%
%
 \date{Received date ; accepted date}
 \maketitle
\markboth{G. J. Qiao et al.: Inverse Compton Scattering model of
pulsar emission: II}{}

\begin{abstract}
The shapes of pulse profiles, especially their variations with
respect to observing frequencies, are very important to understand
emission mechanisms of pulsars, while no previous attempt has been
made in interpreting the complicated phenomenology. In this paper,
we present theoretical simulations for the integrated pulse
profiles and their frequency evolution within the framework of the
inverse Compton scattering (ICS) model proposed by Qiao (1988) and
Qiao \& Lin (1998). Using the phase positions of the pulse
components predicted by the ``beam-frequency figure'' of the ICS
model, we present Gaussian fits to the multi-frequency pulse
profiles for some pulsars. It is shown that the model can
reproduce various types of the frequency evolution behaviors of
pulse profiles observed.
\end{abstract}


\section{Introduction}
A wealth of observational data on radio pulsars has been collected
since their discovery (for a review, see Lyne \& Graham-Smith 1998).
In history, understanding the nature of pulsar radio emission have
been developed along two lines. On the one hand, the characteristics
of pulsar pulse profiles have been studied empirically aiming at an
understanding towards the emission beams. On the other hand,
various plasma instabilities have been
extensively studied motivated to find the right coherent mechanism to
interpret the extremely high brightness temperatures observed. After
more than 30 years, consensus on either of these two issues is yet
fully achieved.

Diverse conclusions are reached after the various
attempts in investigating pulsar radio emission beam shapes.
Rankin (1983; 1993) proposed that the emission
beam is composed of two distinct types of emission components, which
are known as the core emission component near the center and the (two)
conal components surrounding the core. Lyne \& Manchester (1988)
confirmed the different properties of core and conal
emission, but suggested that the observations are better described by a
gradual change in emission characteristics from the core or axial
region to the outer edge emission beam, and that the pulsar
radio beam is patchy, rather than a core plus two cones. The debate
between the ``core-cone'' beam picture and the ``patchy'' beam picture
has been persisting ever since (e.g. Gil \& Krawczyk 1996; Mitra \&
Deshpande 1999; Han \& Manchester 2001; Gangadhara \& Gupta 2001,
among others). It is possible that the real pulsar emission beam is
the convolution of a ``patchy'' source function and a ``window''
function (Manchester 1995), which may itself be composed of a central
core component plus one or more nested ``conal'' components.
Despite of the discrepancy on the pulsar emission beam shapes, a
wealth of multi-frequency observational data has been accumulated
recently (e.g. Kramer et al. 1994; Gould \& Lyne 1998; Kuzmin et
al. 1998). The great varieties of the pulse profiles as well as
their frequency evolution suggest the complexity of the pulsar beam
patterns. For example, in some pulsars some wing (conal) components
emerge at higher frequencies (e.g. PSR 1933+16, Sieber et al 1975),
while in some other pulsars certain components disappear as the
observational frequency evolves (e.g. PSR 1237+25, Phillips \&
Wolszczan 1992). The phase separation between various components also
vary with frequency. All these provide invaluable information about
the pulsar emission beams, and put important constraints on any
theoretical model.

On the theoretical aspect, owing to the extreme environment
within the pulsar inner magnetospheres (strong magnetic fields and
electron-positron plasma), the identification of the pulsar radio
emission mechanism has been a formidable task. More than ten radio
emission models have been proposed. Among these, most models
mainly focus on the condition for developing the instability
which gives rise to coherent emission (Melrose 1992 for a
review). Though theoretically rigorous, most of these models are
either not well-modeled to be compared with the wealth of the
observational data, or obviously conflict with the observations.
The latest discussions include relativistic plasma emission
(Melrose \& Gedalin 1999), plasma maser (Lyutikov, Blandford \&
Machabeli 1999), spark-associated solitons (Gil \& Sendyk 2000;
Melikidze, Gil \& Pataraya 2000), and inverse Compton scattering
(Qiao \& Lin 1998, hereafter Paper 1; Xu et al. 2000). Melrose
(2000) argued that there is now a preferred pulsar emission
mechanism which involves beam-driven Langmuir turbulence, based
on the fact that such an instability involves a frequency
$\nu_{\rm GJ}$ which is $\propto n_{\rm GJ}^{1/2} \propto
B^{1/2}P^{-1/2}$ (see also eq.(2)), and is insensitive to the
environments of either a normal pulsar or a millisecond pulsar.
However, he pointed out a severe difficulty of the mechanism,
i.e., the characteristic emission frequency is far too high
compared to the observed frequency unless some non-standard
plasma condition is introduced (see also Melrose \& Gedalin
1999). Some efforts in comparing the model predictions with the
observational data have been made within the maser model and the
soliton model. However, it remains unclear whether the
above-mentioned variaties of the frequency-evolution of the pulse
profile patterns could be interpreted within these models. As far
as we know, no attempt has been made in understanding the
broad-band pulse profiles within the framework of other models,
except the work by Sieber (1997) how considered a geometrical
effect. The aim of the present paper is to carry out a simulation
of the various frequency evolution patterns of pulsar integrated
pulse profiles over a wide frequency range, within the framework
of the inverse Compton scattering (hereafter ICS) model.

The arrangement of this paper is as follows. In \S2, we review the
basic picture of the ICS model, including a discussion of an
important assumption made in the model.
We then focus on a main theoretical result of the ICS model, i.e., the
so-called ``beam-frequency figure'', on which the later simulations
rely. In \S3, we present some multi-frequency pulse profile
simulations for some typical pulsars of various types, using a
Gaussian-fit according to the pulsar phase predictions made from the
beam-frequency figure, and show how the model can reproduce various
multi-frequency observational data. Our results are summarized in \S4
with some discussions.

\section{The ICS model and the ``beam-frequency figure''}

The basic picture of the ICS model is (Qiao 1988; Paper 1): a
low-frequency electromagnetic wave is assumed to be excited near the
pulsar polar cap region by the periodic breakdown of the inner gap
(Ruderman \& Sutherland 1975), and to propagate
outwards in the open field line area up to some limited heights of
interest. These low energy photons are inverse Compton scattered by
the secondary particles produced in the pair cascades, and the
up-scattered radio photons just provide the observed radio emission
from the pulsar. Recent detailed polar cap ``mapping'' by Deshpande \&
Rankin (1999) indicates that the periodic storm induced by the gap
breakdown is indeed happenning at least in some pulsars, which paves a
solid
observation foundation for the ICS model. It is then natural to expect
the formation of a low frequency electromagnetic wave with the
characteristic frequency $\nu_0 \sim c/h \sim 10^6-10^7 {\rm Hz}$ and
its harmonics, where $c$ is the speed of light and $h$ is the gap
height. The secondary pairs streaming out from the polar cap cascade
with typical energy $\gamma=1/\sqrt{1-\beta^2} \sim 10^3$ will scatter
these low
frequency waves, and the up-scattered frequency reads (for $B\ll
B_q=4.414\times 10^{13}$Gauss)
\begin{equation}
\nu \simeq 2\gamma ^2\nu_0(1-\beta\cos\theta _i),
\label{ics}
\end{equation}
where $\theta_i$ is the incident angle (the angle between the
direction of the particle and the incoming photons) of the
scattering. Given a typical near-surface value of $(1-\beta
\cos\theta_i) \sim 10^{-3}$, one has $\nu \sim 10^9 {\rm Hz}
\gamma_3^2 \nu_{0,6}$, which is the typical frequency of the pulsar
radio emission. This makes ICS a plausible candidate for the pulsar
radio emission mechanism. An important potential problem is the
propagation of the low frequency wave in the pulsar magnetosphere.
It is commonly believed that copious secondary pairs are produced at
the upper boundary of the inner gap with a number density $n=\zeta
n_{\rm GJ}$, where $\zeta \sim 10^3-10^4$ is the multiplicity
(e.g. Daugherty \& Harding 1982), and
$n_{\rm GJ}\sim \Omega B/2\pi ec \sim 7\times 10^{10}~{\rm cm^{-3}}
B_{12}/P$ is the
Goldreich-Julian number density. The plasma frequency of this pair
plasma is then
\begin{equation}
\omega_p=\left({4\pi ne^2 \over m_e\gamma}\right)^{1/2} = 1.5\times
10^{10} \left({\zeta_3 B_{12} \over P\gamma_3}\right)^{1/2} ~{\rm
s^{-1}},
\label{omegap}
\end{equation}
where $B_{12}$ is the local magnetic field strength in unit of
$10^{12}$ G, and $P$ is the rotation period of the pulsar. Near the
surface, this frequency is clearly much higher than $\nu_0$, which
means that the low frequency wave excited by the gap breakdown may not
be able to propagate. In fact, such waves are not detected, so that
they must have been damped within the pulsar magnetosphere. However,
it is still possible that the low frequency wave may propagate within
the inner magnetosphere and be inverse Compton scattered before being
damped. Strictly speaking, the argument that the low-frequency wave
can not propagate only holds when the magnetosphere above the upper
boundary of the gap is filled with the pair plasma. This might be the
situation if the inner gap has a stationary pair formation front,
which is the case of the stable space-charge-limited flow accelerator
(e.g. Arons \& Scharlemann 1979). In the breakdown picture we have
assumed here (Ruderman \& Sutherland 1975; Deshpande \& Rankin 1999),
the pair plasma is ejected in clumps, followed after each ``sparking''
process. This leaves many spaces between the adjacent plasma clouds with
much less dense plasma so that the low frequency wave may propagate as
if in vacuum. The plasma clumps will not spread until reaching the height
\begin{equation}
h_{sp} \sim \gamma^2 h \sim 10^9~{\rm cm} \gamma_3^2 h_3,
\label{hsp}
\end{equation}
which is about 100 times of stellar radii, beyond which the fast
plasma will catch up with slow plasma to form a nearly uniform
plasma. Empirical pulsar theories usually indicate that pulsar
emission region is below or around this altitude, which raises the
possibility of interpreting radio emission in terms of the ICS
mechanism presented here. A full description of the propagation
of the electromagnetic wave in such a rather unsteady, clumpy
plasma is a very difficult task, and was not explored by previous
authors. Here (and also in all the previous papers on the ICS
model) we will assume that the low frequency wave may propagate to
the necessary height of interest (e.g. below $h_{sp}$, eq.(3)),
while postpone a detailed analysis of the propagation problem in
future works. Our aim is to investigate the possible consequence
of this strong assumption, and compare the model results to the
observational data.

\begin{figure*}
\centering\includegraphics[width=6cm]{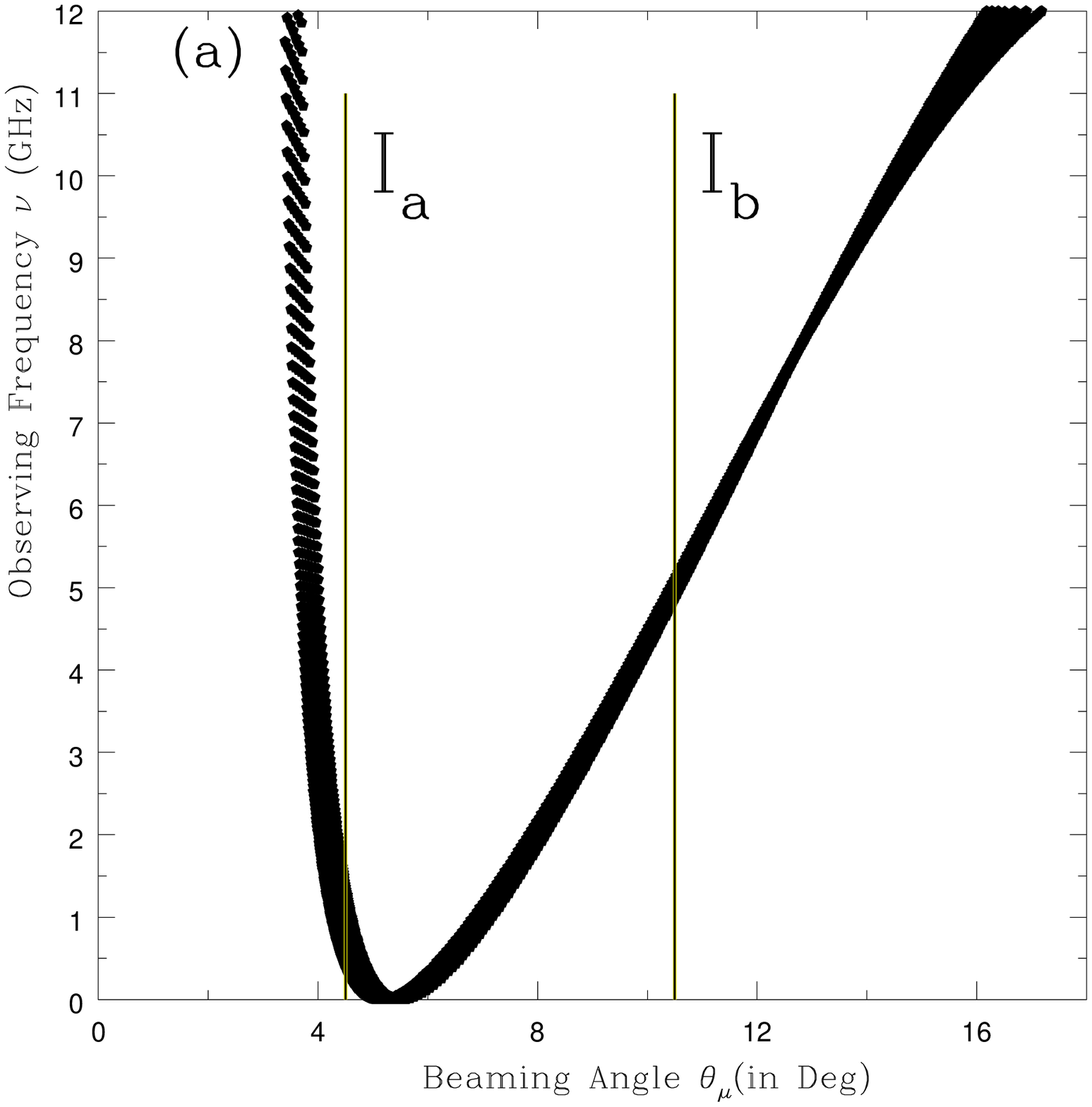}
\includegraphics[width=6cm]{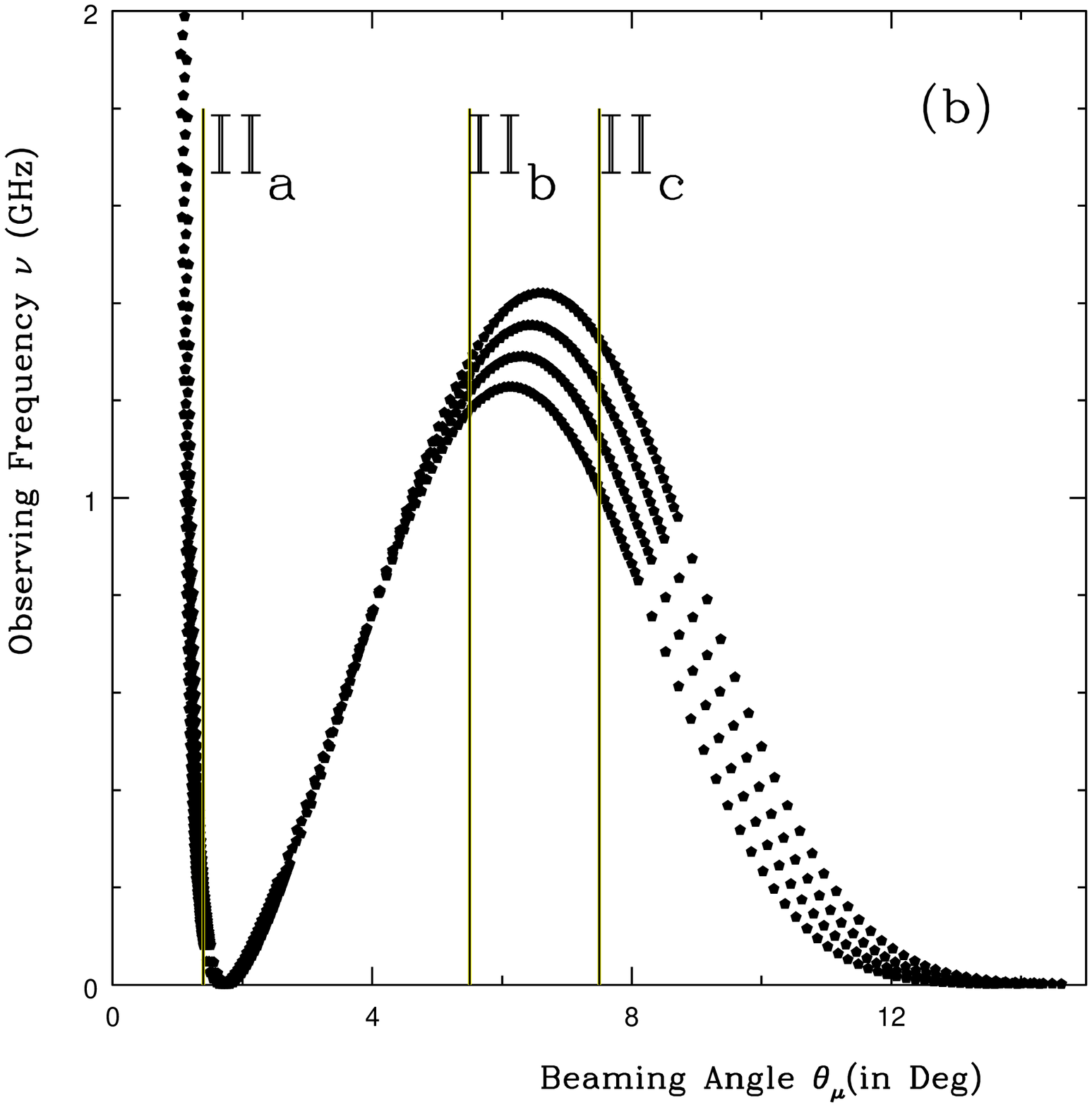}
\caption{Two typical "beam-frequency figures" in the ICS model.
The observing frequency is plotted versus the so-called beaming
angle, the angle between the radiation direction and the magnetic
axis.  (a) For Type I, the lines of a and b correspond to
sub-types Ia and Ib. (b) For type II, the lines of a, b, and c
correspond to the sub-types of IIa, IIb,and IIc.} \label{fig1}
\end{figure*}

If one naively regards that the low frequency wave propagates in
the the near-surface polar cap region as if in vacuum, one can
get an interesting picture of pulsar radio emission beam directly
using eq.(1). Since the secondary pairs usually decelerates due to
various energy loss mechanisms (e.g. Zhang, Qiao \& Han 1997),
one may usually assume
\begin{equation}
\gamma=\gamma_0\exp(-\xi\frac{r-R}{R}),
\end{equation}
where $\gamma_0$ is the initial Lorentz factor of the particles,
$R$ is the radius of of the neutron star, $r-R$ is the emission
height from the neutron star surface, and $\xi$ reflects the
scale factor of the deceleration. On the other hand, basic
dipolar geometry (Fig.1 in Paper 1) indicates that the term
$(1-\beta\cos\theta _i)$ will first decrease with height, then
increase sharply after passing the crossing point defined by
$\theta_i=0$, and finally only increase mildly. The competition
between this geometric effect and the deceleration effect then
results in a drop-rise-drop pattern for the emission frequency as
the height increases. Since the field opening angle increases
with the height, the drop-rise-drop pattern also holds for the
frequency-beaming angle plot. This is the so-called
``beam-frequency figure'' or the ``beaming figure'' of the ICS
model, which is the starting point of the simulations in $\S3$.
Figure 1a,b are typical beam-frequency figures of the ICS model.
Fixing an observation frequency $\nu$, there are three heights
(and opening angles) where the emission with this frequency comes
out. This naturally gives rise to a central core emission
component and two conal components, meeting Rankin's (1983)
original proposal. One remark is that the three-component scheme
is an average picture, which resembles Manchester's (1995)
``window'' function. In a certain pulsar, some favored spots may
generate more clumps and hence, more emission, so that the actual
emission beam could be patchy within the three-component window.

In a beam-frequency figure, we can define the three branches as
the ``core branch'', the ``inner conal branch'' and the ``outer conal
branch'', respectively, as indicated in Fig.1. How many branches are
observable also depends on the line-of-sight of the observer, which
defines the minimum beam angle.

\section{Frequency behaviour of the integrated pulse profiles }

In this section, an important observational feature, i.e., the
integrated pulse shape evolution with frequencies, will be
investigated. The key points are how many components in a pulse
profile exist and what positions of these components are, which
can be retrieved from the ``beam-frequency figures'' as described
above. Once this information is available, we then assume that
the shape of each emission component is Gaussian, as has been
widely adopted in many other studies (e.g. Kramer et al. 1994;
Kuzmin et al. 1996; Wu et al. 1998). Since the height and the
width of Gaussian function are hard to derive from the first
principle, we take them as inputs to meet the observations. We
can then finally get the integrated pulse profiles of a pulsar
for various frequencies. In the following, we will show some
simulation examples for several different types of the pulsar
profiles and their frequency evolution. A preliminary
consideration has been previously presented by Qiao (1992) and
Liu \& Qiao(1999).

\begin{figure*}          
\centering
\includegraphics[width=6cm,height=8cm]{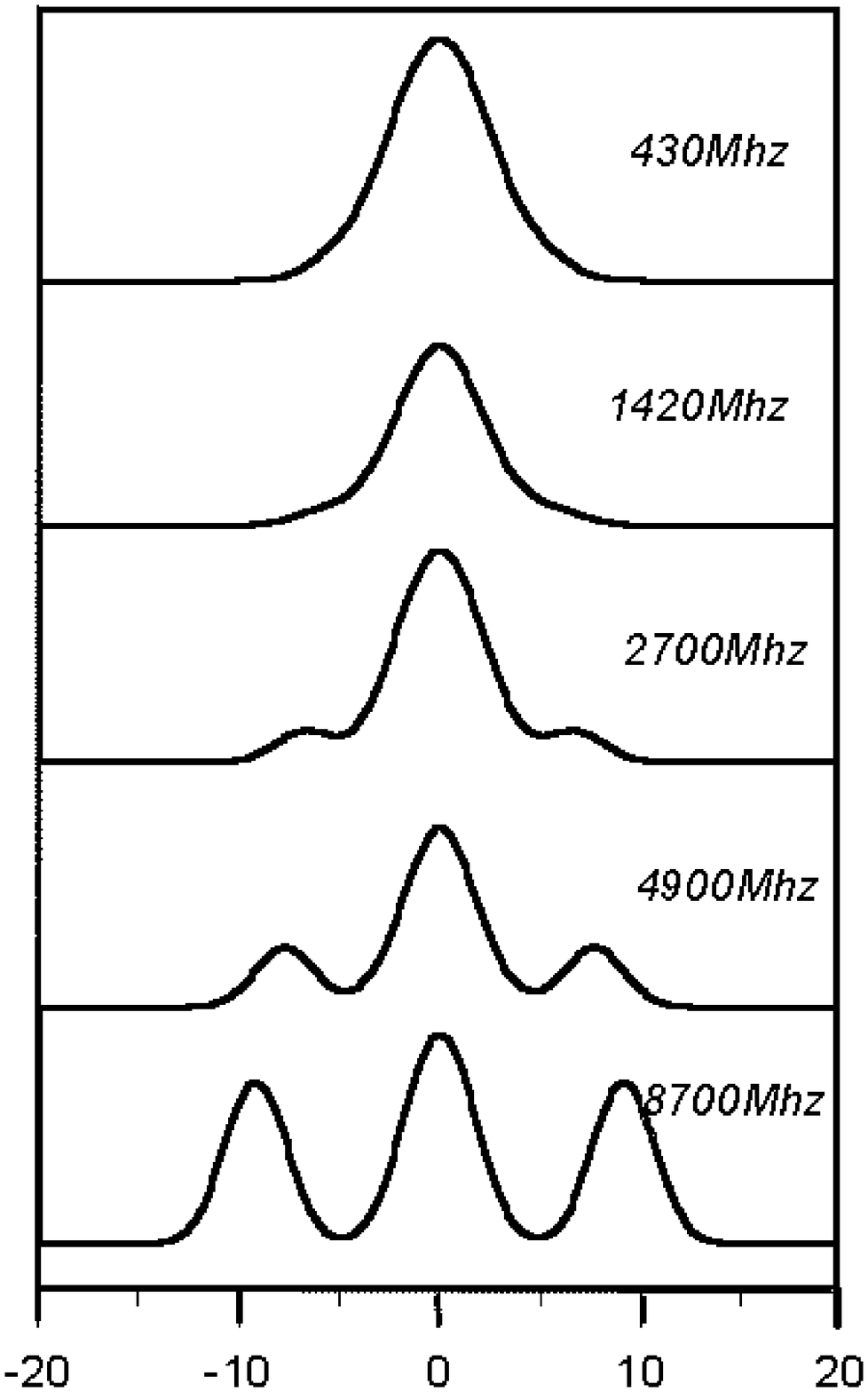}
\includegraphics[width=6cm]{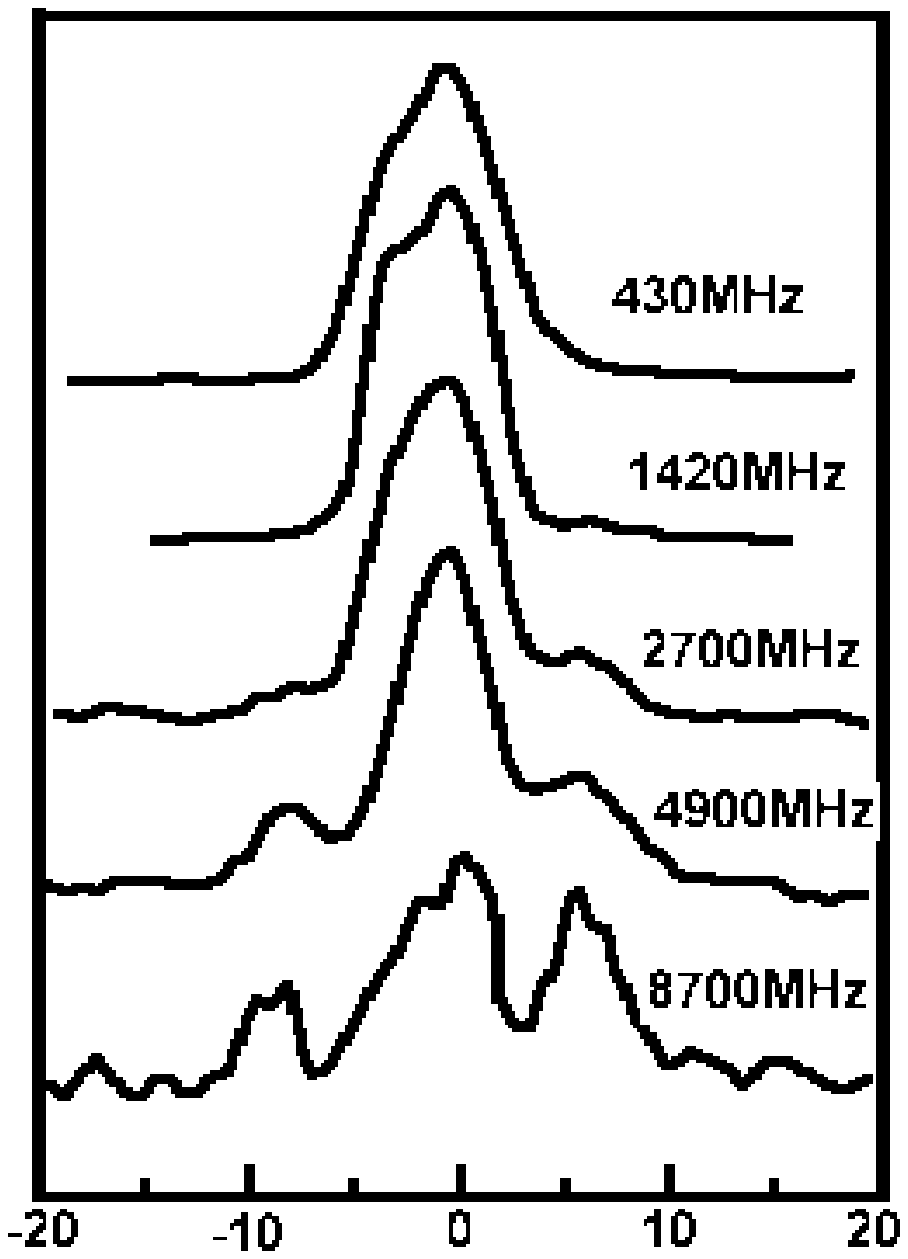}
\caption{A simulation for the type Ia pulsar PSR B1933+16.
$P=0.3587 s$, the inclination angle $\alpha=45^\circ$, the impact
angle $\beta=0$, $\gamma_0=3000$, $\xi=10^{-4}$. We compute the
locations of the cones using the field line whose base is 0.8 of
the polar cap radius. In the right column figure we present the
observational data from Lyne \& Manchester (1988) for 1420MHz,
2700MHz and 8700MHz, and from Sieber et al. (1975) for 430MHz and
4900MHz, for a closer comparison.  Hereafter left columns
represent the simulation results; while right columns are the
multi-frequency observational results. The horizontal axis is the
longitude (in degrees). }
\end{figure*}

\subsection{Core dominant pulsars}

The scheme of this kind of pulsars is that only the core branch
and the inner conal branch exist in the pulsar beam. This usually
takes place in the short period pulsars whose polar caps
are larger (e.g. Fig.4 and Fig.6b in Paper I) than those of the
long period pulsars. The missing of the outer conal branch is either
due to that the deceleration of the pairs is not important or due to
that the low frequency wave may not propagate to higher altitudes.
As the impact angle gradually increases, pulsars of this kind can be
further grouped into two sub-types.

\noindent{\bf Type Ia}. Core-single to core-triple pulsars

Pulsars of this type have very small impact angles. They normally
show single pulse profiles at low frequencies but become triple
profiles at high frequencies when the line-of-sight starts to cut
across the inner conal branch.

The multi-frequency observations of PSR B1933+16 (Sieber et al.
1975; Lyne \& Machester 1988) show that it belongs to this type. Its
profiles are single when observation frequencies are lower than 1.4
GHz, but become triple at higher frequencies. A simulation for PSR
B1933+16 is presented in Fig.2. We can see the single-to-triple
profile evolution with increasing frequency. Furthermore, the
separation between the two ``shoulders'' of the triple profile
gets wider at higher frequencies.  Evidently the radius of the
``inner'' cone increases at higher frequencies, which is an important
feature of the ICS model distinguished from the others (see Sect.3.2
of paper 1).


\begin{figure*}          
\centering
\includegraphics[width=7cm,height=9cm]{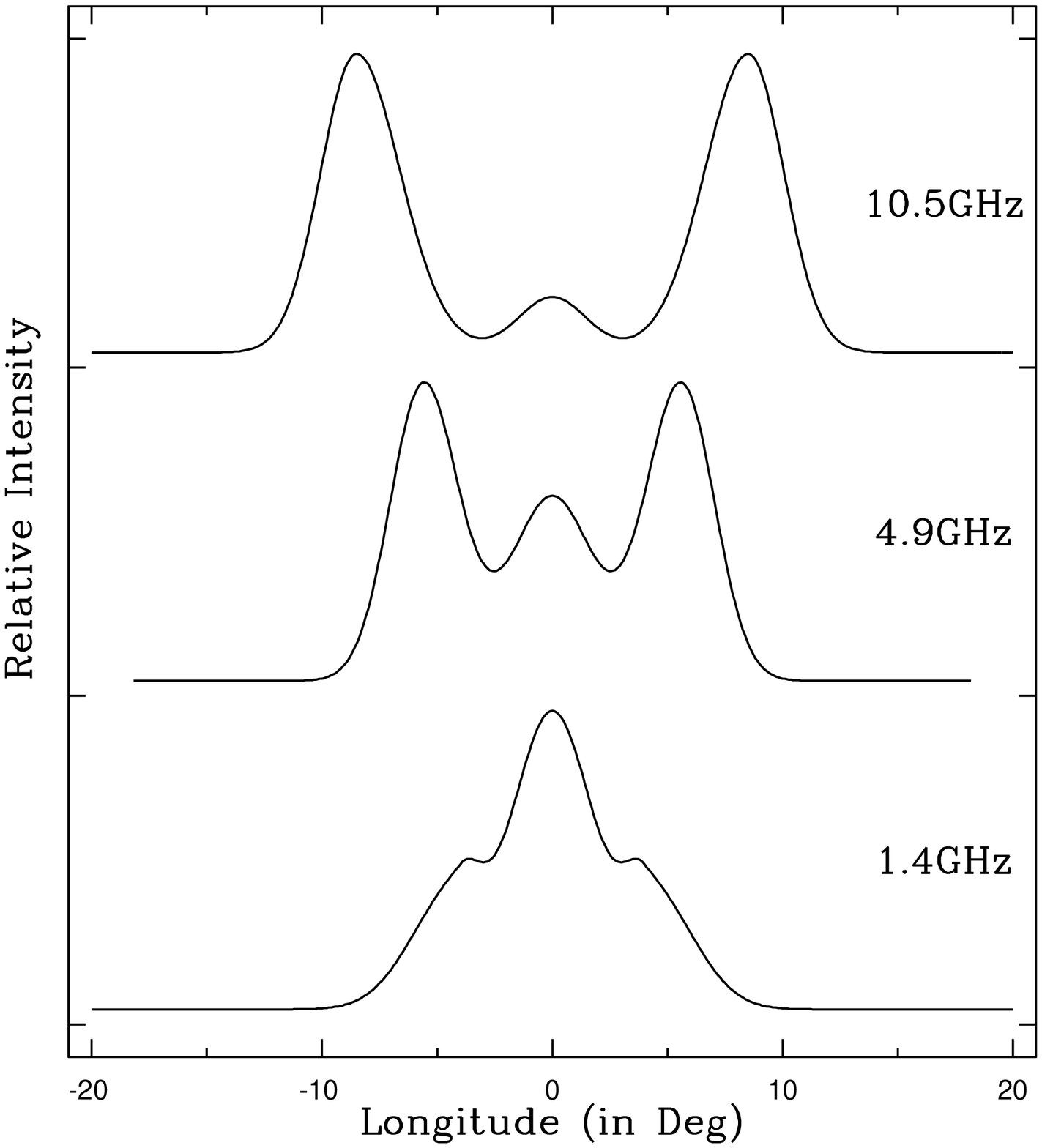}
\includegraphics[width=7cm,height=9.2cm]{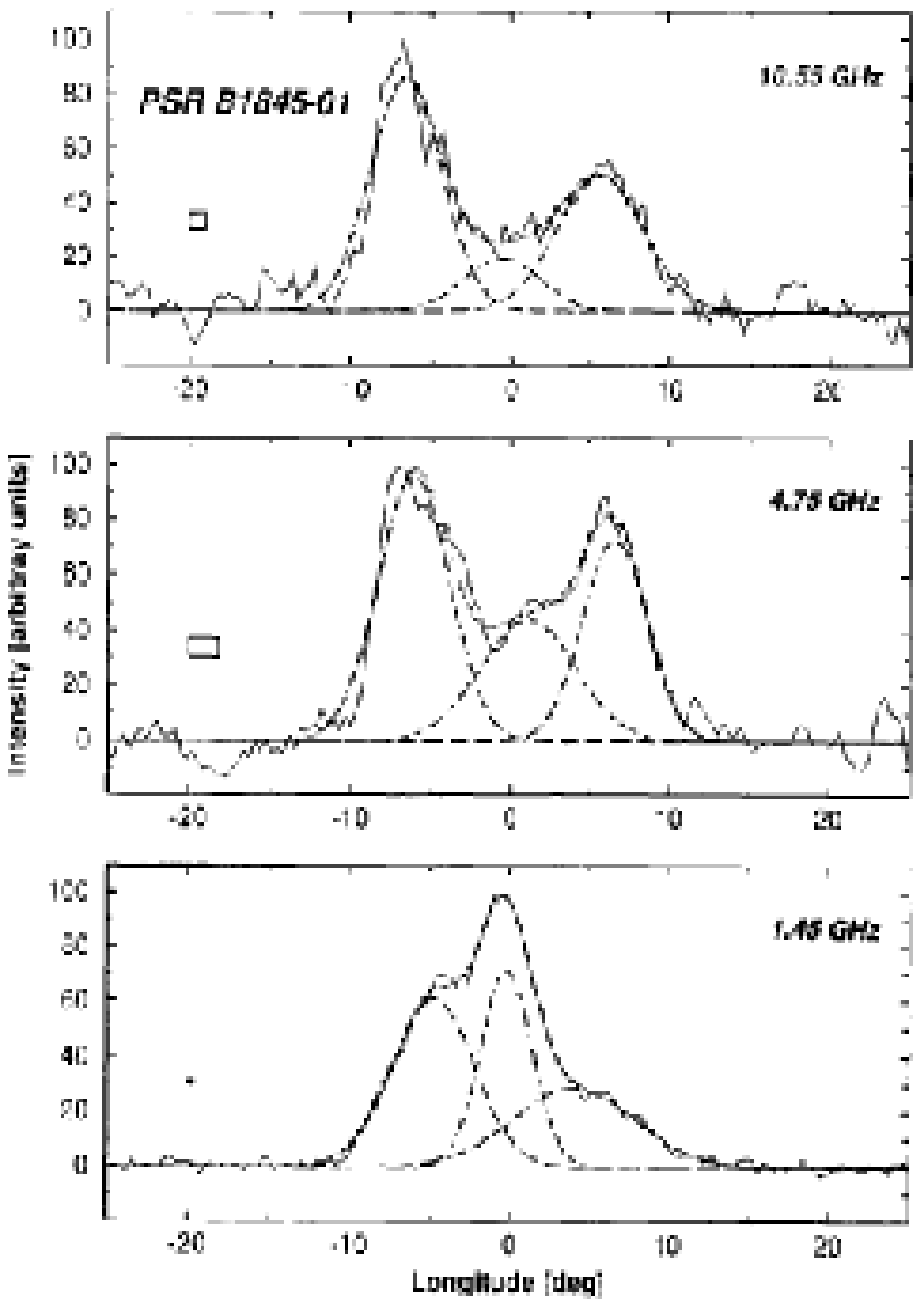}
\caption{A simulation for the Type Ib pulsar PSR B1845-01. P=0.651
s, Model parameters: $\alpha=45^\circ$, $\beta=2.1^\circ$ is large
enough that at higher frequencies the core component is almost
disappearing. $\gamma_0=3000$, $\xi=10^{-4}$. The observations
are from Kramer (1994). }
\end{figure*}


\noindent{\bf Type Ib}. Core-triple to conal-double pulsars

Pulsars of this type have larger impact angles than those of Type
Ia. Though at low frequencies the pulsars show core-triple profiles,
they will evolve to conal-double profiles at higher frequencies, when
the lines-of-sight starts to miss the core branch. An example of this
type is PSR B1845-01 (see Fig.3 and Kramer et al. 1994 for the relevant
data). From both the ICS model and observations one can see several
important characteristics of this kind of pulsars. These are: (1) The
radius of the inner cone increases with the increasing observing
frequency, which is different from that of the outer cone; (2) Owing
to the line-of-sight effect the intensity of the central emission
component decreases rapidly as the observing frequency
increases. These characteristics are very different from those of any
model involving curvature radiation. Another example of this type is
PSR B1508+55 (Kuzmin et al. 1998; Sieber et al. 1975).

\begin{figure*}          
\centering
\includegraphics[width=7cm,height=90mm]{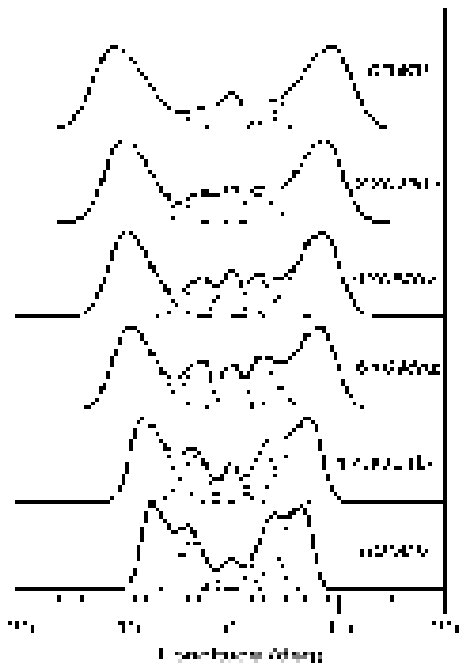}
\includegraphics[width=7cm,height=90mm]{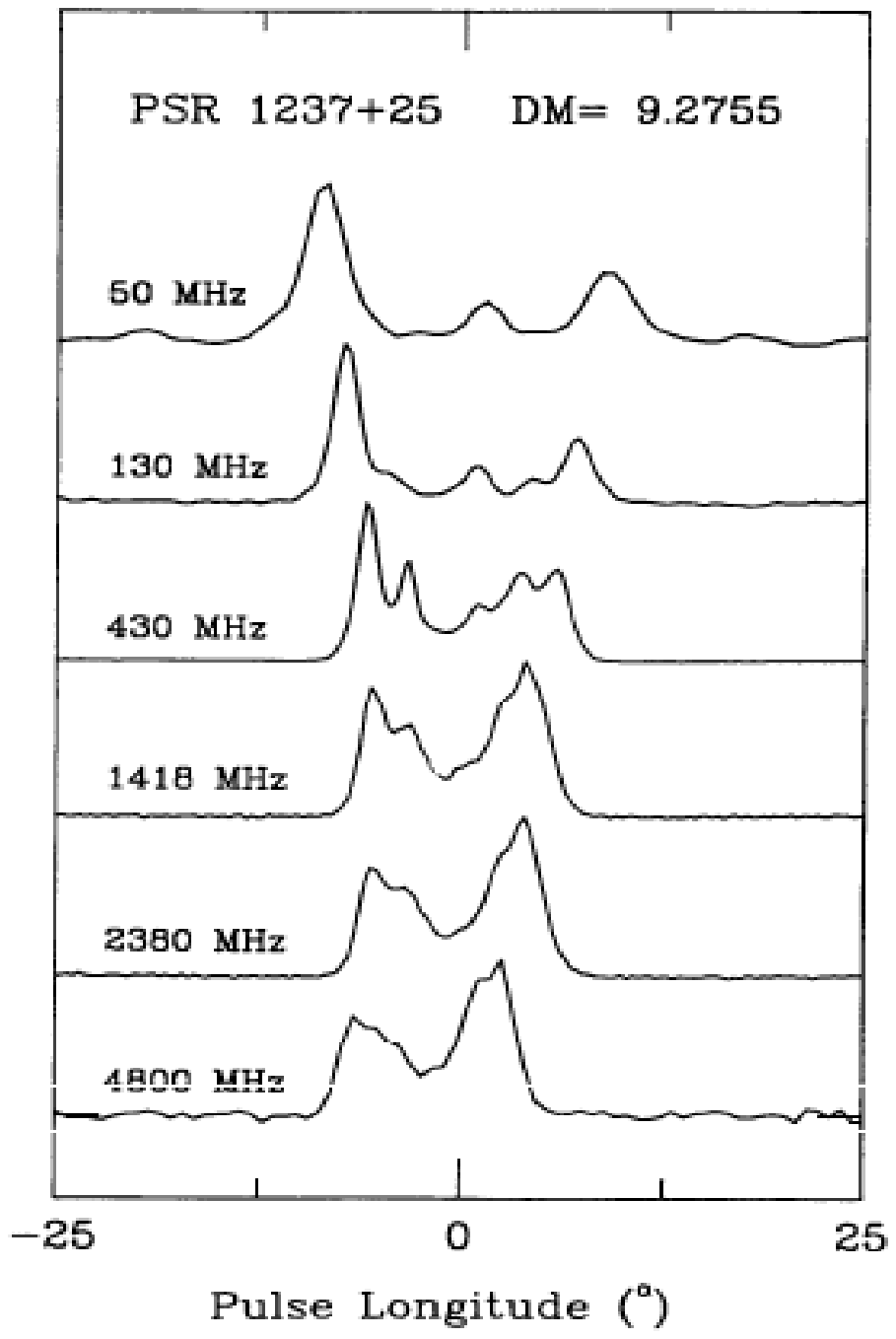}
\caption{A simulation for the Type IIa pulsar PSR B1237+25. P=1.38
s, $\alpha=48.2^\circ$, $\beta=1^\circ$, $\gamma_0=4000$, and
$\xi=0.02$. Note that the retardation effect is taken into
consideration and the central component is off-centered.
Observations are from Phillips \& Wolszczan (1992). }
\end{figure*}

\begin{figure*}          
\centering\includegraphics[width=8cm]{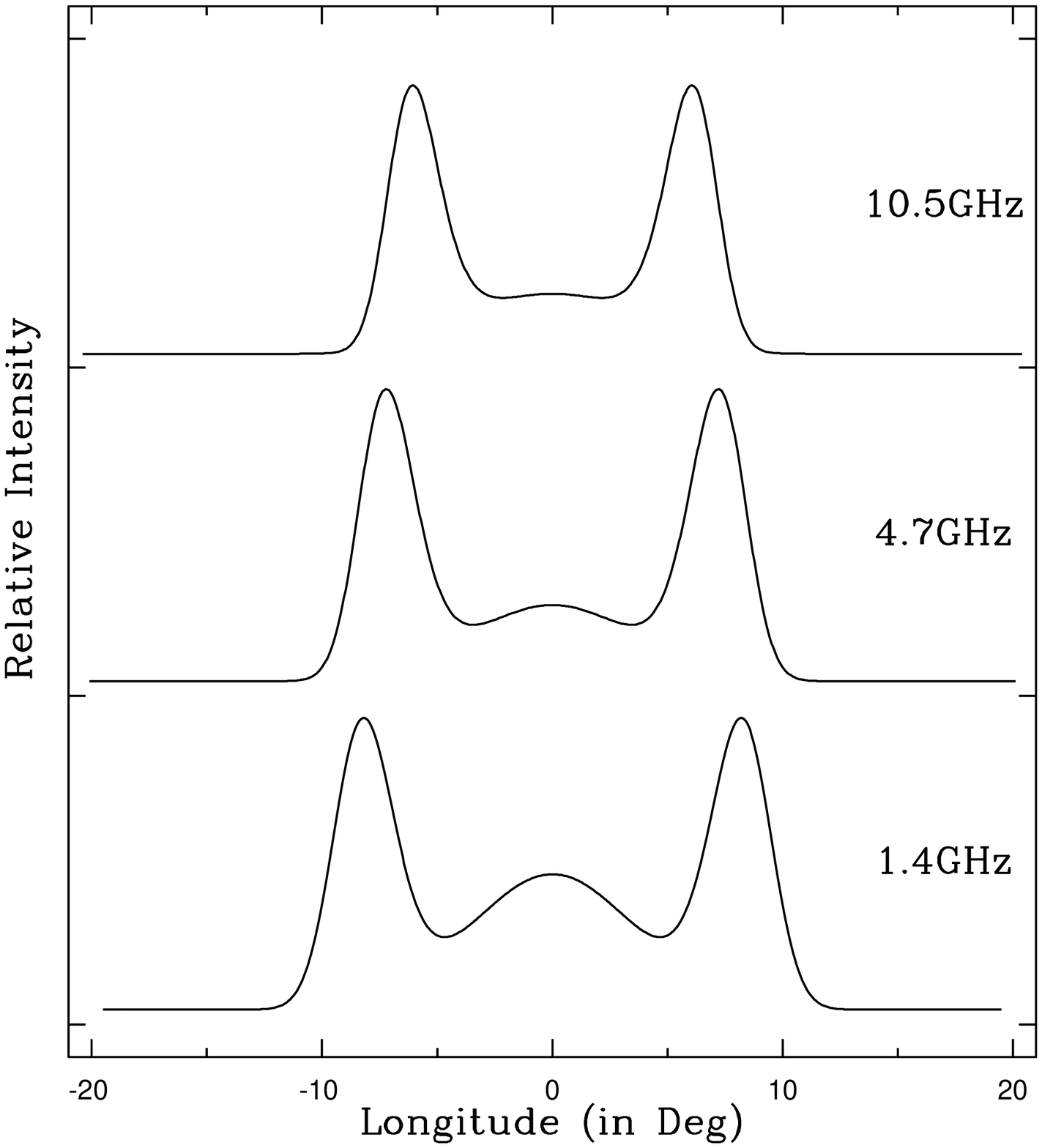}
\includegraphics[width=5.9cm]{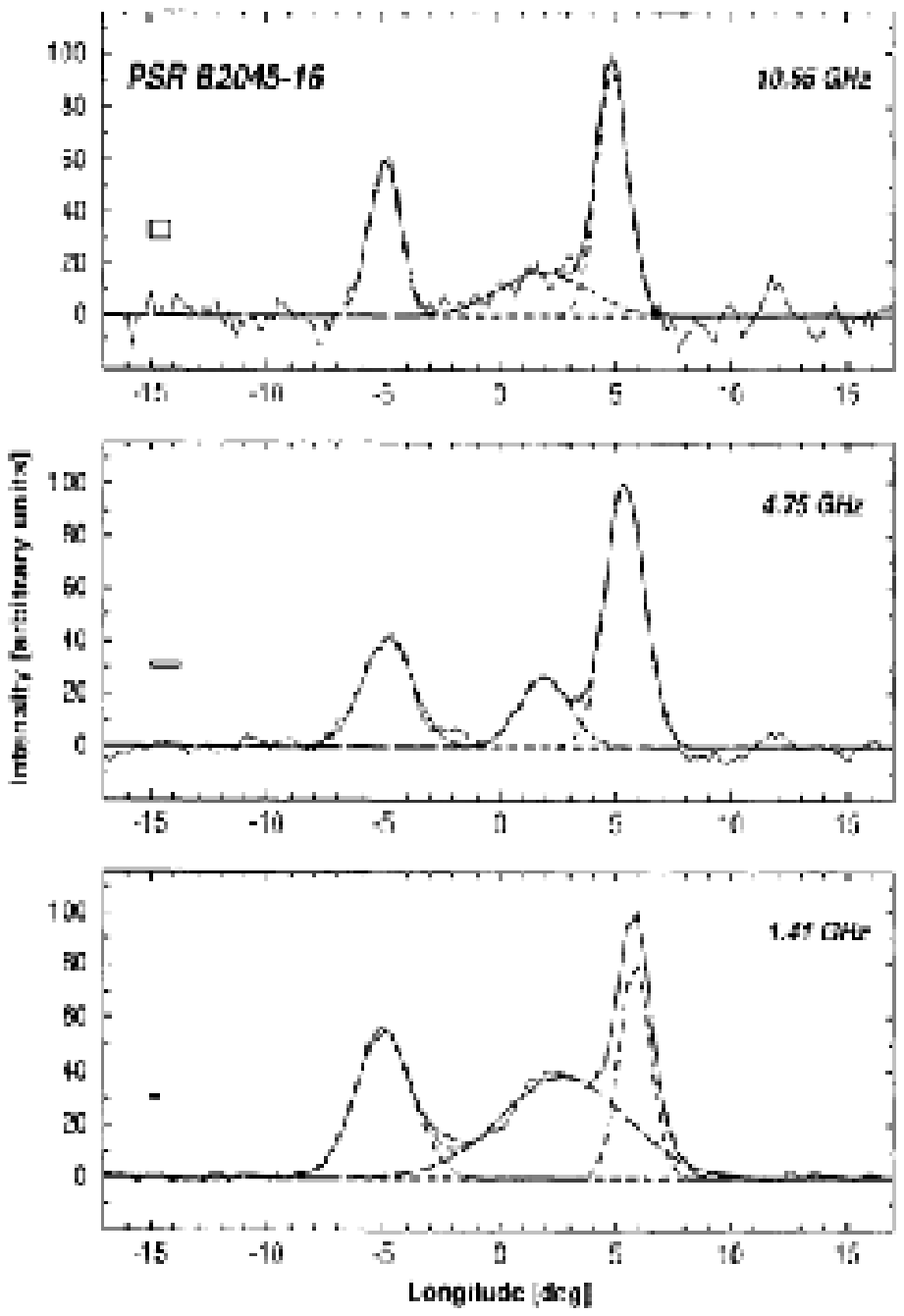}
\caption{A simulation for the Type IIb pulsar PSR B2045-16. P=1.96
s, $\alpha=45^\circ$, $\beta=9^\circ$, $\gamma_0=5000$, and
$\xi=0.002$.   The observations follow Kramer (1994).}
\end{figure*}

\subsection{Conal dominant pulsars}

The scheme of this kind of pulsars is that all three branches
exist in the pulsar beam. This usually occurs in long period pulsars,
mainly because the turning point in the beaming figure ($\theta_i=0$)
gets shifted to lower altitudes due to the geometric effect.
This would be the most common case. Pulsars in this scheme can be
grouped into three sub-types as
the impact angle gradually increases.

\noindent{\bf Type IIa} Multi-component pulsars

Pulsars of this type have five components at most observing
frequencies, since the small impact angle makes the line-of-sight cut
through all the three branches. A very important feature is that
according to the ICS model, the five pulse components will merge into
three components at very low frequencies (see Fig.1b, line IIa in this
paper). This has indeed been observed from some pulsars, e.g., PSR
B1237+25 (see Phillips \& Wolszczan 1992, noticing the three
components at very low frequency of 50 MHz).  It is worth emphasizing
that the ICS model has the ability to interpret this important
feature, which would be a challenge to most of other models. The
simulation results for PSR B1237+25 are presented in Fig.4. Because
three emission components are emitted at different heights, the
retardation is important to perform a realistic simulation. In our
simulation, such a retardation effect was taken into account to
reproduce a peculiar observed feature of this pulsar, i.e., an
off-centered central component closer to the trailing conal
components.

\begin{figure*}          
\centering\includegraphics[width=8.2cm]{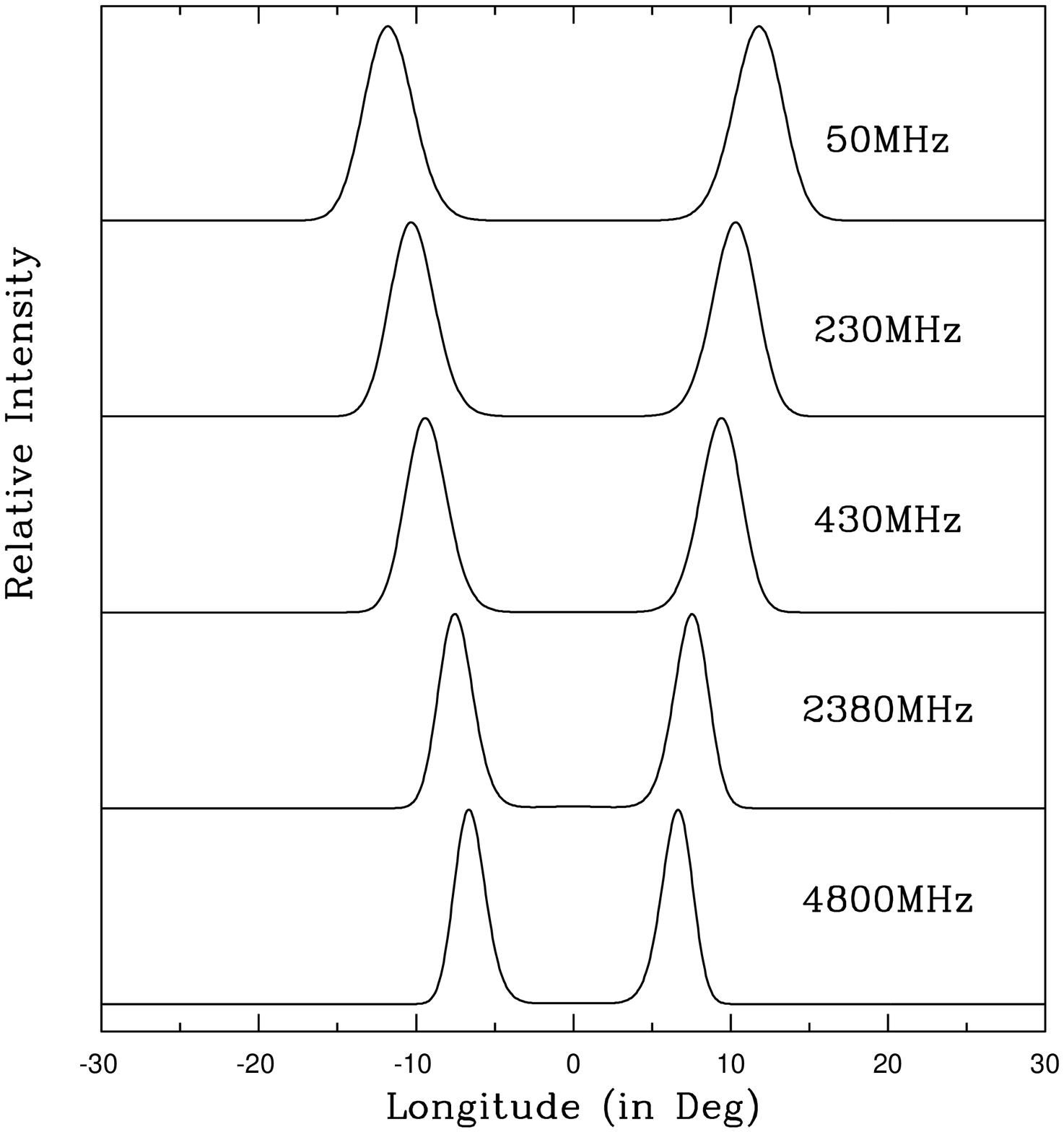}
\includegraphics[width=6cm]{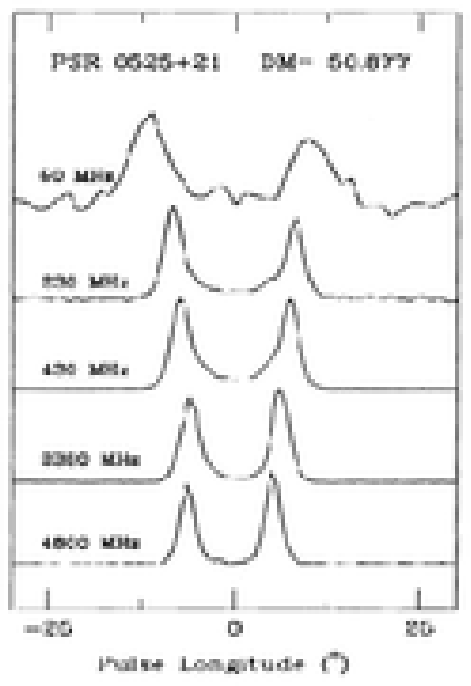} \caption{A
simulation for the Type IIc pulsars PSR B0525+21. P=3.745 s,
$\alpha=45^\circ$, $\beta=6^\circ$, $\gamma_0=5000$, and {\bf
$\xi=0.002$}.  The observations are from Phillips \& Wolszczan
(1992). }
\end{figure*}

\noindent{\bf Type IIb} Conal triple pulsars

The impact angle is larger in this sub-type, so that at higher
frequencies, the line-of-sight does not cut through the core
branch. Thus pulsars show three components at low observing
frequencies, then evolve to four components when frequency is
higher, and finally merge to double pulse profiles at the highest
frequency.

An example is PSR B2045-16 (e.g. Kramer et al. 1994), which we have
simulated in Fig.5. The pulse profile is ``triple'' for
this pulsar. The separation of the two outer components are
wider at lower frequencies.

\noindent{\bf Type IIc} Conal double pulsars

Pulsars of this type have the largest impact angle, so that
only the outer conal branch is cut by the line-of-sight. The
pulse profiles of this type are conal double at all observing
frequencies, with the separation between the two
components decreasing at higher frequencies. This is just the
traditional radius-to-frequency mapping which has also been
calculated in the curvature radiation picture.
A typical one is PSR B0525+21 (Phillips \& Wolszczan 1992), and we
have simulated it in Fig.6.

Another situation of this type is that pulsars have single profiles
at most of observing frequencies, but become double components at
very low observing frequencies, an example of this kind may be
PSR B0950+08 (e.g. Kuzmin et al. 1998).

\section{Conclusions and discussions}

1. Based on a simple ICS picture, we have simulated the frequency
evolution of the pulsar integrated pulse profiles of several pulsars.
It is worth mentioning that the observed evolution behavior is quite
complicated, and there are many different kinds of evolutionary
patterns. This would be a challenge to most presently discussed radio
emission models. In this paper, we find a clear scheme to understand
such a variety of the evolutionary styles within the ICS
model. Different kinds of frequency evolution styles could be grouped
into two basic categories, each of which may be grouped into some
further sub-types according to the line-of-sight effect. Though there
are some uncertainties due to the model assumption, the simulations
presented here can give a first-order description of the pulsar
profiles. The fundamental difference between the ICS and the curvature
radiation mechanism is that the latter can only give hollow
cones. Due to the monotonic beaming-frequency figure of curvature
radiation, formation of the different emission components should be
attributed to the multi-components of the sparking sources (e.g. Gil \&
Sendyk 2000). In the ICS model, one sparking source can naturally
account for three emission components at different altitudes due to
the special beaming figure (Fig.1).

2. As discussed in \S2, we have introduced a strong assumption
throughout the analysis, i.e., the gap sparking-induced low
frequency electromagnetic wave can propagate up to a certain
height and be scattered by the secondary pairs before being
damped at higher altitudes. We have discussed that the strong
unsteady pair process near the pulsar polar cap region may make
this possible, although we concede that more rigorous and
detailed justification about this assumption is desirable.
Nevertheless, the naive picture introduced here seems to have the
ability to reproduce various types of the pulsar integrated pulse
profile patterns and their frequency evolution, which would be
difficult for most of the other theoretically rigorous models.
Another caveat about the propagation problem is that according to
eq.(2), even the observed radio emission can not propagate near
the surface, while observations show that at least some emission
components (e.g. the core emission) may come from the surface
(Rankin 1990). Some other ideas that may allow propagation of the
low frequency wave include the radiation-pressure-induced plasma
rarefication (Sincell \& Coppi 1996) and the non-linear plasma
effect (e.g. Chian \& Kennel 1983).

3. Observationally, it has been argued that different emission
components may come from different heights (e.g. Rankin 1990;
1993). More specifically, Rankin (1990) found that the profile widths
of the core single pulsars are remarkably consistent with the
prediction of a near-surface emission with dipolar field
configuration, which hints that the core emission may come from the
near polar cap region. If it is indeed so, then the ICS model gives a
natural explanation of the near-surface core emission. All the other
presently discussed models (Melrose 2000; Lyutikov et al. 1999;
Melikidze et al. 2000) exclusively predict a much higher emission
altitude for both the core and the conal emission components.

4. Our model calculations show that in order for the low
frequency wave to be generated, pulsars should have oscillatory
inner gaps. This is a natural expectation if the inner gap of a
pulsar is vacuum-like (Ruderman \& Sutherland 1975) rather than a
steady space-charge-limited flow (Arons \& Scharlemann 1979). In
the conventional neutron star picture, this requires that the
star is an ``anti-parallel rotator'', i.e., $\Omega\cdot B < 0$,
with a large enough ion binding energy on the surface. If pulsars
are born with random orientations of spin and magnetic axes, the
present model would then only applies to one half of the neutron
stars. If the space-charge-limited accelerator can somehow show
an oscillatory behavior (e.g. Muslimov \& Harding 1997), then our
model can also apply to the other half of the neutron stars.
Furthermore, if pulsars are strange quark stars with bare polar
cap surfaces (e.g. Xu, Qiao \& Zhang 1999), one expects a vacuum
gap forming in both $ \Omega\cdot B < 0$ and $ \Omega\cdot B > 0$
configurations. Another caveat is that there might be more than
one radio emission mechanisms operating in pulsars. For example,
some young pulsars have an emission component with almost 100\%
linear polarization, very likely with a high-altitude wide cone
configuration. This component may be due to some other reasons
(e.g. the maser model by Lyutikov et al. 1999), and may be also
produced by a steady space-charge-limited accelerator. Future
broadband observations (radio, optical, X-ray, and $\gamma$-ray)
from some pulsars may eventually reveal the geometric
configurations for the emission components in various bands, and
shed some lights on our understanding of the pulsar radio
emission mechanisms.

\acknowledgements {We are grateful to the referees for their
careful reviews and the suggestions that helped to improve the
presentation of the paper, and to Profs. R. N. Manchester, J. M.
Rankin and J. A. Gil for their discussions about the ICS model.
We also thank many useful discussions with the members in our
group, Drs. Xu, R.X., Hong, B.H., and Mr. Wang, H.G., and
especially, thank Wang H.G. and Xu,R.X for valuable technical
assistence. This work is partly supported by NSF of China, the
Climbing project, the National Key Basic Research Science
Foundation of China, and the Research Fund for the Doctoral
Program Higher Education.}


\end{document}